\documentclass[twocolumn,showpacs,preprintnumbers,amsmath,amssymb]{revtex4}
\input epsf.sty

\usepackage{graphicx}
\usepackage{dcolumn}
\usepackage{bm}

\begin{document}


\title{Precise estimation of shell model energy by
second order extrapolation method
}

\author{Takahiro Mizusaki}
\affiliation{Institute of Natural Sciences, Senshu University, Higashimita, 
Tama, Kawasaki, Kanagawa, 214-8580, Japan}

\author{Masatoshi Imada}%
\affiliation{%
Institute for Solid State Physics, University of Tokyo,  
Kashiwanoha, Kashiwa, 277-8581, Japan
}%

\date{\today}

\begin{abstract}
A second order extrapolation method is presented for shell model
calculations, where shell model energies of truncated spaces
are well described as a function of energy variance by quadratic curves
and exact shell model energies can be obtained by the extrapolation.
This new extrapolation can give more precise energy  than those
of first order extrapolation method.
It is also clarified that first order extrapolation gives a lower limit of 
shell model energy. 
In addition to the energy, we derive the second order extrapolation formula 
for expectation values of other observables.  
\end{abstract}

\maketitle

A pursuit for new numerical algorithms to solve nuclear shell model problems 
has been one of the most intriguing issues in nuclear structure physics.
In the last decade, several new methods \cite{smmc,dmrg,vampir,qmcd1,qmcd-rev,
txp,ecm,xp} were proposed and have been developed. Consequently
shell model calculation remarkably widens its capability 
beyond the limitation of conventional approaches such as Lanczos shell model 
diagonalization. 
Among these new methods, several extrapolation techniques\cite{txp,ecm,xp} were proposed
in order to reinforce a restricted power of Lanczos shell model calculations.
We advocated the Lanczos shell model diagonalization 
with the help of the recently proposed non-empirical extrapolation technique \cite{imada1,sorella} 
for correlated electron systems
and succeeded in enlarging the feasibility of large-scale shell model calculations 
by the extrapolation \cite{xp}. 

In general, interpolation can be reliable if the considered quantity behaves
smoothly, while extrapolation cannot be guaranteed.
A reliable extrapolation method needs a theory predicting how the considered 
quantity behaves in the extrapolated region. 
In the recently proposed extrapolation \cite{imada1,sorella}, it has been shown that, 
exact energies defined in the whole space can be extrapolated from 
series of energies in truncated Hilbert spaces.
If these truncated Hilbert spaces give fairly good approximations, 
energies of truncated spaces can be shown to behave linearly as a function of the
energy variance. Therefore exact energies can be estimated
by linearly extrapolated energies of truncated spaces  into zero energy variance.
In condensed matter physics there are good examples of this extrapolation
in analyses of strongly correlated electrons on a lattice \cite{imada2,imada3}.
In Ref. \cite{xp} it has been shown that this extrapolation
works also when applied to the nuclear shell model.
However, for detailed spectroscopic studies, precise excitation energies are important,
and this requires a more precise evaluation of the energy eigenvalues.
Another problem 
is, as it will be discussed later,  that the extrapolated shell model energies are systematically lower than the 
true energy eigenvalues.
To solve this problem, in the present paper, we successfully improve the extrapolation 
method by including the second order term. Hereafter this extension is called
second order extrapolation, while the previous one is called first order
extrapolation.

In order to present our new  method, we consider the relation between
energy difference and modified energy variance as below.
We define the difference $\delta E$ between the lowest energy eigenvalue 
$\left\langle  {\hat H}  \right\rangle$ in a given truncated  
space and true energy eigenvalue $\left\langle {\hat H} \right\rangle _0$ as
\begin{equation}
\delta E = \left\langle {\hat H} \right\rangle -\left\langle {\hat H} \right\rangle_0.
\end{equation}
The modified energy variance  $\Delta E$ in the truncated space is also defined as
\begin{equation}
\Delta E={{\left\langle {\hat H^2} 
\right\rangle -\left\langle {\hat H} \right\rangle ^2}.
}
\end{equation}
Note that this definition of energy variance 
is different from the previous one \cite{xp}. 
Throughout the present paper, this new definition is used. 

An approximate ground state $\left| \psi  \right\rangle $ can be
decomposed as the sum of the true eigenstate $\left| {\psi _0} \right\rangle$ and 
the rest component $\left| {\psi _r} \right\rangle$ as
\begin{equation}
\left| \psi  \right\rangle =c \left| {\psi _0} \right\rangle 
                           +d \left| {\psi _r } \right\rangle,
\end{equation}
\begin{equation}
  \left| {\psi _r} \right\rangle = \sum\limits_{n\ne 0}
 {c_n\left| {\psi _n} \right\rangle },
\end{equation}
where $c$, $d$ and $c_n$'s are coefficients, whose normalization condition 
is taken as
$c^2+d^2=1$ and $\sum\limits_{n\ne 0} {c_n^2=1}$. 
The  $ \left| {\psi _0} \right\rangle $ and 
$\left| {\psi _n} \right\rangle$'s form an  orthonormarized complete set spanning  
the whole space and its energy eigenvalue is denoted by $E_n$.
By defining the moments $D_j$ ($j=1,2,$...) as 
\begin{equation}
D_j=\sum\limits_{n\ne 0} {{{c_n^2\left( {E_n-E_0} \right)^j} \over {E_0^j}}},
\end{equation}
we can rewrite $\delta E$ and $\Delta E$ as
\begin{equation}
\delta E=d^2D_1E_0,
\end{equation}
and
\begin{equation}
\Delta E=d^2D_2E_0^2-\left( d^2 \right)^2 \left( {D_1E_0} \right)^2.
\end{equation}
By eliminating $d^2$, we obtain 
\begin{equation}
\delta E={1 \over {2A}}\left[ {1-\sqrt {1-4A^2\Delta E}} \right],
\end{equation}
where 
\begin{equation}
A={{D_1} \over {D_2E_0}}.
\end{equation}
If $4A^2 \Delta E$ is small enough compared to the unity, 
we can expand $\delta E$ as
\begin{equation}
\delta E=A\Delta E+A^3\Delta E^2+ ... .
\end{equation}
As the ratio of the moments $A$ embodies the structure of the rest component, $A$ can be 
a function of $d^2$ which is a small parameter. 
Here we assume that $A$ can be expanded by a function of $d^2$ as
\begin{equation}
A=A_0+A'_1d^2+O(d^4),
\end{equation}
where $A_0$ and $A'_1$ are coefficients. This expansion 
is shown to be reasonable in the later discussion.
By eliminating $d^2$ by Eq. (7), $A$ can be rewritten as
\begin{equation}
A=A_0+A_1\Delta E+ ... ,
\end{equation}
where $A_1$ is a new coefficient for $A'_1$ with some additional factors.
Therefore, we obtain the following second order extrapolation 
formula,
\begin{equation}
\delta E=A_0\Delta E+\left( {A_0^3+A_1} \right)\Delta E^2+ ... .
\end{equation}
The expansion parameter of Eqs. (12) and (13) is $\Delta E/E_0^2$
but, as $E_0$ is unknown, we present a new extrapolation formula
by including the $E_0$ depedence in the expansion coefficients.  
We comment that, a dimensionless definition of energy variance as
$\Tilde \Delta E=\Delta E/ {\left\langle \hat  H \right\rangle _0^2}$ 
leads to the same result.

The first term of Eq. (13) shows the effect of the decrease of the norm of rest 
component.
If we consider only the first term, this extrapolation is nothing but the previous 
first order extrapolation \cite{xp}.
The second term shows the second order effects of the decrease of the norm of
rest component ($A_0^3$)  and the first order effect of the structure change ($A_1$).
The sign of the second order coefficient will be clarified later in the discussion.
This new method allows us to extrapolate shell model energies in the limit
$\Delta E\to 0$ after determining the expansion coefficients in Eq. (13)  
by  fitting to the shell model energies of various truncation spaces.

In order to investigate the properties of the present second order
extrapolation method, we utilize numerically solvable shell model calculations.
One frontier of the state-of-the-art large-scale shell model calculations 
is the $fp$ shell. 
Here we take ground state energies of $^{48}$Cr and $^{49}$Cr as $fp$ shell nuclei.
For a numerical test, we consider two cases of different nature.
The $M$-scheme
dimensions are about 2 and 5 Millions for $^{48,49}$Cr, respectively, 
and we can easily diagonalize the Hamiltonian matrices by a shell model code, 
for instance, MSHELL \cite{mshell}.
As an effective shell model interaction, the KB3 interaction \cite{kb3} is taken. 

We consider two kinds of truncation scheme. 
The first is $\oplus_{s\le t}(f_{7/2})^{A-40-s}(r)^s$ 
where $r$ means the set of the $f_{5/2}$, $p_{3/2}$ and $p_{1/2}$ orbits and $t$ 
is the maximum number of particles allowed to be excited into the $r$ orbits. 
This truncation scheme is natural due to a considerable shell gap 
between the $f_{7/2}$ orbit and the others. 
We call this truncation scheme (I).
The other is the truncation scheme (II):
$\oplus_{s+s'\le t_0, s'\le t}(f_{7/2})^{A-40-s-s'}(p_{3/2})^{s}(r')^{s'}$ 
where $r'$ means the set of the $f_{5/2}$ and $p_{1/2}$ orbits.
Truncation spaces are specified by  $t$ and $t_0$. For $^{48,49}$Cr,
the $t_0$ is set to 5, which means up to 5 nucleons in the $p_{3/2}$ and $r'$ orbits.
The parameter $t$ controls the size of the shell model space by specifying the maximum 
number of nucleons in the $r'$ orbits. 
These two sets of truncation scheme  gradually cover the whole shell model space
in different ways as $t$ increases.

\begin{figure}[h]
\begin{picture}(230,225)
    \put(0,0){\epsfxsize 240pt \epsfbox{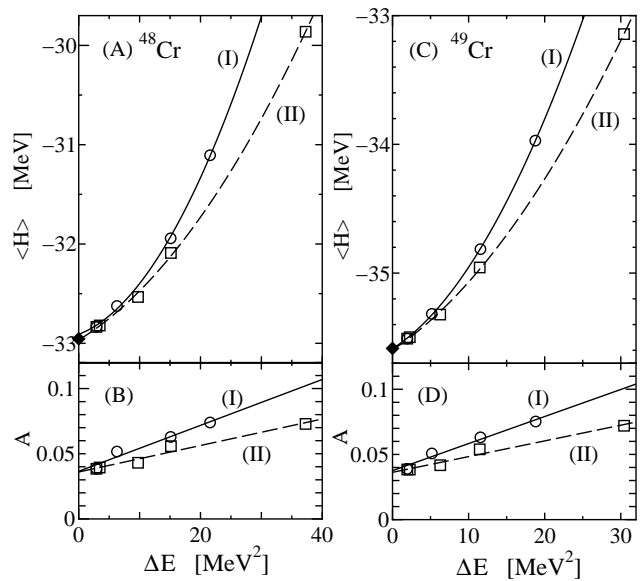}}
\end{picture}
\caption{
Second order extrapolations of ground state energies into 
zero energy variance;  
(A) for ground state ($0^+_1$) of $^{48}$Cr,
and (C) for ground state ($(5/2)^-_1$) of $^{49}$Cr.
Open circles  and open squares are energies for the truncated shell model spaces.
Exact ground state energies are shown by filled diamonds.
Results of two kinds of truncation scheme are shown 
for solid (I) and dashed (II) curves for (A) and (C).
The ratio of the monents $A$ is shown as a function of $\Delta E$
for (B) and (D).
}
\end{figure}

In Figs. 1 (A) and (C), the energies for the truncated shell model spaces 
with  $t=2 \sim 5$ for the truncation scheme (I) and $t=1 \sim 5$ for the 
truncation scheme (II) are shown as a function of the energy variance 
$\Delta E$ for the ground states of $^{48}$Cr and $^{49}$Cr.
The solid (I) and dashed (II) curves show the second order extrapolations.
Independently from the nature of the nuclei and from the adopted truncation scheme,
the shell model energies for truncated spaces are well-fitted by quadratic curves
as indicated by the second order extrapolation formula.
These extrapolated energies are quite accurate, for examples,
for $^{48}$Cr, they are  -32.91 and -32.97 MeV for (I) and (II), 
respectively, while exact energy is -32.95 MeV.
For $^{49}$Cr, extrapolated energies are -35.59 and -35.60 MeV for (I) and (II), 
respectively, while exact one is -35.59 MeV.
Different truncation schemes can give almost the same extrapolation energy.
This fact is quite useful for confirming the reliability of the extrapolated energy.

\begin{figure}[h]
\begin{picture}(230,130)
    \put(0,0){\epsfxsize 180pt \epsfbox{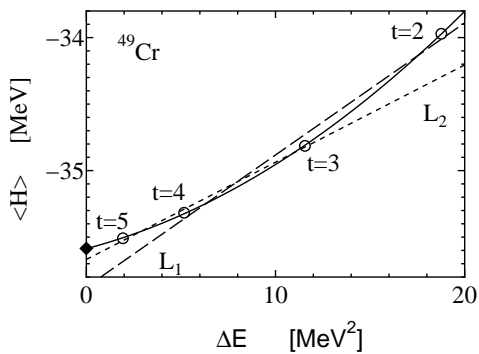}}
\end{picture}
\caption{
First order extrapolations (L$_1$ and L$_2$) and second order extrapolation
(solid curve) are shown for the ground state of $^{49}$Cr.
}
\end{figure}

In Fig.2, we compare the results of the first order extrapolation 
and of the second order extrapolation
for the ground state of $^{49}$Cr, where the truncation scheme (I) is taken.
L$_1$ and L$_2$ are the extrapolations from the shell model energies
for $t=2,3,4$ and $t=3,4,5$, respectively.
L$_2$ gives higher extrapolation energy than L$_1$, while
it is still lower than the exact energy.
However, globally the shell model energies for truncated 
spaces can be  better fitted by a quadratic curve 
than by straight lines.
The first order extrapolation is approximately parallel
to the  average tangential line of 
this quadratic curve around the fitted shell model energies.
If the quadratic curve is upwardly concave, that is  $A_0^3+A_1 >0$,
the extrapolated energy becomes a lower limit of shell model energy and approaches
the exact energy from the lower side. 

We further go into the details of the second order extrapolation formula.
As we can obtain exact ground state wavefunctions for $^{48,49}$Cr,
the moments $D_1$ and $D_2$ can also be evaluated.
In Figs. 1 (B) and (D), the ratio of the moments $A$ is plotted as a function 
of $\Delta E$.
These figures show that $A$ can be well described by the linear 
relation of $\Delta E$.
The coefficient $A_0$ is almost the same for different truncation schemes,
while $A_1$ is different.
For $^{48}$Cr, the moment ratios are
 $A=0.037+0.0018\Delta E$ for (I) and $A=0.036+0.0010\Delta E$ 
for (II).
In the second order coefficient, $A_1$  is positive for both cases.
Moreover,  $A_1$ is much larger than $A_0^3$. 
This means that it is important to take into account the structure change of the 
rest component in second order term.

As the moments $D_1$ and $D_2$ are positive by definition,
 the ratio of the moments $A$ is positive then
$A_0$ is also positive. The sign of the second order coeficient of Eq. (13) depends on
$A_1$. Here we explain the condition $A_1 > 0$.
We define the energy of the rest component relative to the true ground state
energy, $E_r=\left\langle {\psi _r} \right|H\left| {\psi _r} \right\rangle -E_0$.
The first moment $D_1$ is proportional to $E_r$, while
the second moment $D_2$ is  approximately proportional to $E_r^2$. 
Therefore their ratio $A$ has a $1/E_r$ dependence.
If $E_r$ increases as $t$ increases, $A$ decreases as a function of $t$,
which means that $A_1 > 0$.
In the following we consider two  physically relevant but different
truncation schemes that,
however, share a common way for expanding  shell model spaces 
as $t$ increases. The shell model orbits are divided into  higher and lower orbits.
As the parameter $t$ is the maximum nucleon number in the higher orbits, we expand  
truncated spaces by including basis states with larger
spherical single particle energies on average. By this construnction of the 
series of truncation spaces,
as $t$ increases, the rest component is described by basis states with larger 
spherical single particle energies and $E_r$ becomes larger.
Therefore upwardly concave feature is generally expected in energy - energy 
variance plot.

Next we consider the expectation value of an observable $\hat O$,
where a similar second order extrapolation 
formula  holds. In the same way as the energy, we define the difference as
\begin{equation}
\delta O = \left\langle {\hat O} \right\rangle -\left\langle {\hat O} \right\rangle_0,
\end{equation}
that can be rewritten by \cite{imada2}
\begin{equation}
\delta O \sim d^2D_1^O E_0,
\end{equation}
in which  $D_1^O $ is defined as
\begin{equation}
 D_1^O=\sum\limits_{n,m\ne 0} {{{c_n c_m\left( {O_{nm}-O_{00}\delta_{nm}} \right)} 
\over {E_0}}},
\end{equation}
where $O_{nm}= \left\langle  {\psi _m} \left| {\hat O} \right| {\psi _n} \right\rangle$.
By eliminating $d^2$,
$\delta O$ can be expanded by the ratio $\Delta E/E_0^2$ as
\begin{equation}
\delta O=B\Delta E+\left( {B A^2} \right)\Delta E^2+ ... ,
\end{equation}
with
\begin{equation}
B={{D_1^O} \over {D_2E_0}}.
\end{equation}
We assume that $B$ can be expanded as 
\begin{equation}
B=B_0+B_1\Delta E+ ... ,
\end{equation}
where $B_0$ and $B_1$ are coefficients. Then,
we obtain the second order formula for  $\delta O$ as
\begin{equation}
\delta O=B_0\Delta E+\left( {B_0 A_0 ^2+B_1} \right)\Delta E^2+ ... .
\end{equation}

As a further large-scale numerical test, we investigate the ground state 
of $^{64}$Zn, whose $M$-scheme dimension is about 501 Million and 
its exact shell model calculation is quite difficult.
However, exact results are reported in Ref. \cite{caurier-fp}.
 
\begin{figure}[h]
\begin{picture}(230,330)
    \put(0,0){\epsfxsize 170pt \epsfbox{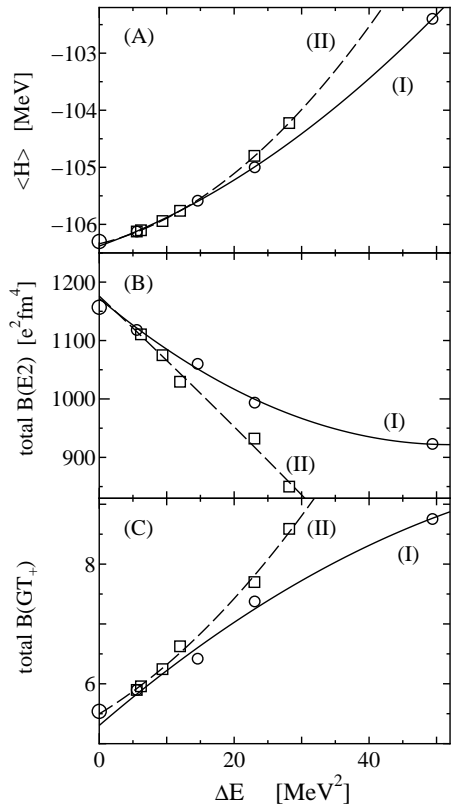}}
\end{picture}
\caption{
(A) Second order extrapolations for the energy to the zero energy variance 
for the $0 ^{+}_{1}$  state of $^{64}$Zn. 
(B, C) Extrapolations of the $B(E2)$ (B) and Gamow-Teller (C)
sum rules for the $0^{+}_{1}$ state 
into zero energy variance.  Exact results \protect\cite{caurier-fp}
are shown by open circles.
The effective charges are taken as  $e_p=1.35e$ and $e_n=0.35e$.
For the oscillator length, $b=1.01A^{1/6}$ fm is taken.
Unquenched Gamow-Teller operator are used.
The solid  and dashed lines show the results of truncation schemes (I)
and (II), respectively.
}
\end{figure}

In Fig. 3, the results of extrapolations of energies (A), $B(E2)$ (B) and Gamow-Teller (C)
sum rules for the $0^{+}_{1}$ state are shown for $^{64}$Zn.
We consider two truncation schemes. For the truncation scheme (I),
we plot the results for $t=1$ $\sim$ 4 spaces, whose $M$-scheme dimension
is 0.2 $\sim$ 55 Millions.
For the truncation scheme (II), we plot the results for $t=4$ $\sim$ 8 with $t_0=4$,
whose $M$-scheme dimension is 0.2 $\sim$ 40 Millions.
Shell model energies of truncated spaces 
can be well-fitted by the quadratic curves for both the truncation schemes. 
Extrapolated energies are -106.38 and  -106.34 MeV, for the truncation schemes (I) and
(II), respectively.
The exact ground state energy is -106.3 MeV \cite{caurier-fp}.
The present extrapolated energies are quite close to the exact one.
The extrapolated values of the total B(E2) are 
1.17 $\times 10^3$ and 1.18 $\times 10^3$ $e^2fm^4$ for the truncation schemes (I)
and (II),
respectively, while the exact value is 1157  $e^2fm^4$ \cite{caurier-fp} with
the effective charges as $e_{\pi}=1.35e$ and $e_{\nu}=0.35e$.
Here we use the second order extrapolation formula for the expectation values.
The extrapolated values of the total B(GT+) with unquenched Gamow-Teller operators
are 5.3 and  5.5 for the truncation schemes (I) and (II), respectively,
while its exact value is 5.54. 
The present extrapolations with two truncation schemes can give almost exact values.  

In summary, we have proposed a second order extrapolation method for shell model calculations.
Shell model energies of the truncated spaces as a function of $\Delta E$ are
well described by a quadratic curve and the exact shell model energy is extrapolated 
in the limit $\Delta E \to 0$.
To derive this second order extrapolation formula, it is quite important to consider the
second order term of the $d^2$ expansion and the structure change 
of rest components in terms of the ratio of the moments $A$.
By numerical solvable shell model calculations, 
we have compared  the extrapolated energies to the exact one 
and showed that their accuracy  is improved
by the present second order extrapolation
and that the first order extrapolation gives a lower limit of
the shell model energy.
In addition, we have extended  
the derivation of second order extrapolation to any observable.
Finally, we have shown that different truncation schemes have different behavior in extrapolation,
but, in principle, extrapolated values must be the same. 
This fact is used to confirm the accuracy of extrapolated values.

We thank Professor  N. Yoshinaga and Dr. M. Serra for reading the manuscript 
and delightful comments. 
This work was supported from 'Research for the Future Program' by Japan Society 
for the Promotion of Science under grant RFTF97P01103, and was supported in part 
by Grant-in-Aid for Scientific Research (A)(2) (10304019), and one for Specially 
Promoted Research (13002001) from the Ministry of Education, Science and Culture.

\end{document}